\global\def\draftcontrol{0}

   \def\versionno{ diffusion -- draft   }

\catcode`\@=11

\expandafter\ifx\csname draftcontrol\endcsname\relax\global\def\draftcontrol{0}
\fi

{\count255=\time\divide\count255 by 60
\xdef\hourmin{\number\count255}
\multiply\count255 by-60\advance\count255 by\time
\xdef\hourmin{\hourmin:\ifnum\count255<10 0\fi\the\count255}}
\def\draftdate{\number\month/\number\day/\number\year\ \ \ \hourmin }

\newcommand\makepapertitle{\par
  \begingroup
    \renewcommand\thefootnote{\@fnsymbol\c@footnote}%
    \def\@makefnmark{\rlap{\@textsuperscript{\normalfont\@thefnmark}}}%
    \long\def\@makefntext##1{\parindent 1em\noindent
            \hb@xt@1.8em{%
                \hss\@textsuperscript{\normalfont\@thefnmark}}##1}%
     \newpage
     \global\@topnum\z@   
     \@makepapertitle
     \thispagestyle{empty}\@thanks
  \endgroup
  \setcounter{footnote}{0}%
  \global\let\thanks\relax
  \global\let\makepapertitle\relax
  \global\let\@makepapertitle\relax
  \global\let\@thanks\@empty
  \global\let\@author\@empty
  \global\let\@date\@empty
  \global\let\@title\@empty
  \global\let\title\relax
  \global\let\author\relax
  \global\let\date\relax
  \global\let\and\relax
  \def\version{\let\version\@version\@gobble}
}
\def\@makepapertitle{%
  \newpage
   \ifnum\draftcontrol=1 {}
   \version\versionno
   \vskip 3em%
   \else
   \hfill\hbox to 3cm {\parbox{4cm}{\@pubnum}\hss}%
   \vskip 3em%
   \fi
   \begin{center}%
   \let \footnote \thanks
     {\LARGE {\@title}}%
     \vskip 1.5em%
     {\normalsize
       \lineskip .5em%
       \begin{tabular}[t]{c}%
         \@author
       \end{tabular}\par}%
     \vskip 1.5em%
     {\@bstract}%
     \end{center}%
     \vskip 1.5em
     \@date%
   \par
}

\gdef\@pubnum{}
\def\pubnum#1{%
  \gdef\@pubnum{#1}}

\gdef\@bstract{}
\def\Abstract#1{%
  \gdef\@bstract{%
   \parbox{\textwidth-0pc}{%
   \centerline{\bf Abstract}\penalty1000%
\kern.2cm%
\noindent
\renewcommand\baselinestretch{1.0}%
{#1}}}
}

\def\ps@paper{\let\@mkboth\@gobbletwo%
     \ifnum\draftcontrol=1
	\def\@oddfoot{\hbox to \textwidth{\tiny \versionno \hfil\tiny\draftdate}%
	\hskip -\textwidth \hbox to \textwidth{\hfil\rm\thepage\hfil}}%
     \else\def\@oddfoot{\hbox to \textwidth{\hfil\rm\thepage\hfil}}
     \fi
     \let\@evenfoot\@oddfoot
}

\def\body{\clearpage
          \pagestyle{paper}
	}

\def\@version#1{\ifnum\draftcontrol=1
\typeout{}\typeout{#1}\typeout{}
\vskip3mm\centerline{\hbox{\fbox{\normalsize{\tt DRAFT -- #1 -- }
                   {\draftdate}}}}\vskip3mm
\fi}
\let\version\@version
\long\def\eqlabel#1{\ifnum\draftcontrol=1
                    \tag@false  
                    \tag*{(\theequation) \hbox to -0.2cm{\hspace{0cm}\small{#1}\hss}}
                    \refstepcounter{equation}
                    \edef\@currentlabel{\theequation}
                    \ltx@label{#1}          
                    \else
                    \label{#1}
                    \fi
                    }
\let\st@bibitem\@bibitem
\let\st@lbibitem\@lbibitem
\ifnum\draftcontrol=1
  \def\@bibitem#1{%
    \st@bibitem{#1}\a@@label{#1}\ignorespaces}
  \def\@lbibitem[#1]#2{%
    \st@lbibitem[#1]{#2}\a@@label{#2}\ignorespaces}
  \def\a@@label#1{%
    \gdef\a@lab{\smash{\normalfont\small#1}}
    \ifvmode
      \if@inlabel
        \global\setbox\@labels\hbox{%
          \llap{\a@lab\let\a@lab\relax
                \kern\@totalleftmargin\kern\marginparsep}%
          \box\@labels}%
      \fi
    \fi}
\fi

\documentclass[12pt,letterpaper]{article}

\usepackage{amsmath,amssymb,array,calc,rotating,epsfig,psfrag}
\usepackage[nosort]{cite}

\ifnum\draftcontrol=1
\fi

\renewcommand\baselinestretch{1.25}
\renewcommand\section{\@startsection {section}{1}{\z@}%
                                   {-3.5ex \@plus -1ex \@minus -.2ex}%
                                   {2.3ex \@plus.2ex}%
                                   {\normalfont\large\bfseries}}
\renewcommand\subsection{\@startsection{subsection}{2}{\z@}%
                                   {-3.25ex\@plus -1ex \@minus -.2ex}%
                                   {1.5ex \@plus .2ex}%
                                   {\normalfont\normalsize\bfseries}}
\renewcommand\subsubsection{\@startsection{subsubsection}{3}{\z@}%
                                   {-3.25ex\@plus -1ex \@minus -.2ex}%
                                   {1.5ex \@plus .2ex}%
                                   {\normalfont\normalsize\it}}
\renewcommand\paragraph{\@startsection{paragraph}{4}{\z@}%
                                   {-3.25ex\@plus -1ex \@minus -.2ex}%
                                   {1.5ex \@plus .2ex}%
                                   {\normalfont\normalsize\bf}}

\numberwithin{equation}{section}

\def\ie{{\it i.e.}}

\def\revise#1       {\raisebox{-0em}{\rule{3pt}{1em}}%
                     \marginpar{\raisebox{.5em}{\vrule width3pt\
                     \vrule width0pt height 0pt depth0.5em
                     \hbox to 0cm{\hspace{0cm}{%
                     \parbox[t]{4em}{\raggedright\footnotesize{#1}}}\hss}}}}

\def\cald         {{\cal D}}

\def\caln         {{\cal N}}
\def\calo         {{\cal O}}

\def\tr           {\mathop{\rm Tr}}

\def\sqr#1#2{{\vcenter{\vbox{\hrule height.#2pt
 \hbox{\vrule width.#2pt height#1pt \kern#1pt
 \vrule width.#2pt}\hrule height.#2pt}}}}

\newcommand{\fft}[2]{{\frac{#1}{#2}}}

\def\om{\Omega}

\def\a{\alpha}
\def\b{\beta}

\def\dd{\delta}
\def\tg{\tilde{g}}

\def\n{\nabla}

\def\bD3{\overline{D3}}

\def\b{\beta}
\def\P{\Phi}
\def\t{\triangle}
\def\hg{\hat{g}}
\def\tg{\tilde{g}}
\def\hr{\hat{r}}
\def\tr{\tilde{r}}

\catcode`\@=12

\begin{document}


\title{Universality of the shear viscosity in supergravity}

\pubnum{%
NSF-KITP-03-103\\
MCTP-03-53\\
hep-th/0311175}
\date{November 2003}

\author{
Alex Buchel$^{1,2}$ and James T. Liu$^{3}$\\[0.4cm]
\it $^1$Perimeter Institute for Theoretical Physics\\
\it Waterloo, Ontario N2J 2W9, Canada\\[0.1cm]
\it $^2$Department of Applied Mathematics\\
\it University of Western Ontario\\
\it London, Ontario N6A 5B7, Canada\\[0.1cm]
\it $^3$Michigan Center for Theoretical Physics\\
\it Randall Laboratory of Physics, The University of Michigan\\
\it Ann Arbor, MI 48109-1120  
}

\Abstract{
Kovtun, Son and Starinets proposed a bound on the shear viscosity 
of any fluid in terms of its entropy density. We argue that 
this bound is always saturated for gauge theories at large 
't~Hooft coupling, which admit holographically dual
supergravity description. 
}


\makepapertitle

\body

\version\versionno

\section{Introduction}

One of the remarkable connections arising from holography has been the
link between black hole thermodynamics and the more traditional case on
the field theory side.  By working with a black hole (or black brane)
background on the gravity side, this allows the investigation of the
corresponding gauge theory at finite temperature.  Such a connection
alone has already yielded many new insights on the thermal phase structure
of gauge theories.  On the other hand, it is important to realize that
basic equilibrium thermodynamic quantities, such as the free energy and
entropy do not provide complete information about the theory.  In
principle, with an exact AdS/CFT prescription, it ought to be possible
to provide dual descriptions of any desired process in the gauge theory.

In practice, of course, we do not expect to find a simple description
encompassing all of the information of the gauge theory.  However, in
keeping with thermodynamic ideas, it is natural to expect that the
long-distance fluctuations in the theory will have a hydrodynamic
description.  In this manner, one may expand the study of gauge theories
at finite temperature to encompass, {\it e.g.}, transport phenomenon
such as diffusion and sound propagation
\cite{h1,h2,h3,Policastro:2002tn,Herzog:2003ke}.  Along these lines, Kovtun,
Son and Starinets (KSS) \cite{kss} extended the previous results of
\cite{h1,h2,h3} and investigated the shear viscosity, $\eta$, for a large
variety of backgrounds.  

In the examples of \cite{kss}, which cover all maximally supersymmetric
gauge theories and $\caln=2^*$ gauge theory (to leading order in $m/T$)
\cite{pw,n2}, it was found that the ratio of shear viscosity $\eta$ to
the entropy density $s$ had a fixed value, $\eta/s=1/4\pi$.  On the
other hand, it can be shown that
coupled systems have $\eta/s\gg1$, and even common substances have
ratios well about this value, which upon reintroducing fundamental constants 
becomes
\begin{equation}
\frac{1}{4\pi}\to \frac{\hbar}{4\pi k_B}\approx 6.08\ \times\ 10^{-13}\
{\rm K\cdot s}\,.
\eqlabel{hk}
\end{equation}
Based on these observations, KSS  conjectured that there is a universal
bound in nature for this ratio, namely \cite{kss}
\begin{equation}
\frac{\eta}{s}\ge \frac{1}{4\pi}\,.
\eqlabel{bound}
\end{equation}
It was further argued in \cite{k} that this bound follows from the
generalized covariant entropy bound \cite{bousso}.  

The intriguing result that $\eta/s=1/4\pi$ holds exactly for many different
nonextremal brane backgrounds suggests that saturation of the bound
\eqref{bound} may always be true for systems admitting a dual supergravity
realization.  In this letter, we demonstrate that this is in fact always
the case.  Before doing this however, we provide a brief review of the
hydrodynamics of strongly coupled systems in section~2.  Then in section~3
we point out that the bound \eqref{bound} is saturated for $\caln=2^*$
gauge theory \cite{pw}, Klebanov-Tseytlin (KT) gauge theory \cite{kt},  and 
Maldacena-Nunez (MN) gauge theory \cite{mn}.  This leads us, at the end of
the section, to prove that the bound is always saturated in the supergravity
approximation of strongly coupled gauge theories. 

Of course, as nature has demonstrated, the shear viscosity bound is not
necessarily saturated at weak coupling.  This, however, is not in
contradiction with our proof, as one cannot directly compare strong and
weak coupling results.  This was already been seen in a different context
in {\it e.g.} the case of the entropy of ${\cal N}=4$ super-Yang-Mills
theory at weak and strong coupling.  Nevertheless, the conjecture that
there is a minimum value of $\eta/s$ still remains to be tested.  In
particular, in order to verify the validity of \eqref{bound} in the
framework of gauge/string correspondence, one has to go beyond the
supergravity approximation and include $\a'$ corrections (corresponding
to finite 't~Hooft coupling corrections).  We consider some aspects of
this issue in section~4.

\section{Hydrodynamics and diffusion in supergravity duals}

Just as in thermodynamics, hydrodynamics is not concerned with the
microscopic properties of a theory, but instead in its macroscopic
ones.  Overall, hydrodynamics may be invoked to provide an effective
description of long wavelength and long time properties of a
macroscopic medium.  In this section, we briefly review a few key
ideas in the application of hydrodynamics to the study of strongly
coupled gauge theories.

Of particular interest to hydrodynamics is the study of diffusion
governing the flow of say heat or charge through a medium.  Assuming
we have a charge related to a conserved current, the diffusion of the
charge is then governed by its local concentration, so that
$\vec j=-D\vec\nabla j^0$.  Combining this with current conservation (also
thought of as the continuity equation), $\partial_t j^0+\vec\nabla\cdot\vec
j=0$, then yields the familiar heat equation, $\partial_t
j^0=\vec\nabla\cdot(D\vec\nabla j^0)$.  As expected for a thermodynamic
description, these equations are no longer Lorentz invariant, and time
reversal invariance is explicitly broken.

While this is all well known, its application to AdS/CFT is perhaps
less familiar.  The important point here is that the diffusion
coefficient $\cald$ is connected to the underlying properties of the
gauge theory.  At the same time, techniques have been developed to extract
$\cald$ from the fluctuations of long-wavelength modes in the supergravity
dual \cite{h1,h2,h3,kss}.  So for strongly coupled gauge theories where
the dual is known, computation of $\cald$ and other kinetic coefficients yields
additional insight on the nature of the theory itself.
                                                                                
Of more direct concern to us is bulk transport through a medium.  In this
case, one works with energy, momentum and pressure, or in other words a
conserved stress-energy
tensor with components $T^{00}$, $T^{0i}$ and $T^{ij}$.  While the analysis
is similar to that of charge diffusion, some additional complication
arises from the tensor nature of $T^{\mu\nu}$.  The resulting hydrodynamic
quantities of interest include the bulk viscosity $\zeta$, shear viscosity
$\eta$, and the speed of sound $v_s$.

In order to compute these kinetic coefficients from the gravity dual,
one may in principle extract the appropriate behavior of the boundary
stress tensor $T^{\mu\nu}$.  Alternatively, as demonstrated in \cite{kss},
the shear viscosity may be determined by setting up a `shear perturbation'
as a fluctuation on top of the original supergravity background, given
by the metric
\begin{equation}
ds^2 = (G_{00}(u)dX_0^2+G_{xx}(u)d\vec x\,^2)+G_{uu}(u)du^2+\cdots\,,
\end{equation}
where the dual gauge theory has $(X_0,\vec x\,)$ coordinates, $u$ is the
transverse coordinate, and the ellipses denote compact directions which
are not of direct concern in the following.  This metric has a
plane-symmetric horizon that extends in $p$ infinite spatial directions,
$\vec x=\{x^i\}$. We assume this metric has a horizon at $u\to u_0$ where
$G_{00}$ vanishes.  The decay of
this shear mode is then governed by a diffusion coefficient
\begin{equation}
\cald=\frac{\sqrt{-G(u_0)}}{\sqrt{-G_{00}(u_0)G_{uu}(u_0)}}\int_{u_0}^{\infty}
du\ \frac{-G_{00}G_{uu}}{G_{xx}\sqrt{-G}}\,,
\eqlabel{diff}
\end{equation}
denoted the shear mode diffusion constant in \cite{kss}.

The shear viscosity, $\eta$, may be extracted from the diffusion
constant ${\cal D}$ according to \cite{kss}
\begin{equation}
\cald=\frac{\eta}{\epsilon+P}=\frac 1T\ \frac{\eta}{s}\,,
\eqlabel{did}
\end{equation}
in the dual gauge theory.  Here $\epsilon$, $s$, $P$ and $T$ are
correspondingly the equilibrium energy and entropy densities, the pressure
and temperature.
As a result, the conjectured shear viscosity bound, \eqref{bound}, is
equivalent to the statement
\begin{equation}
{\cal D}\ge\fft1{4\pi T}\,.
\end{equation}
This is the form of the bound that we use in the subsequent sections.

\section{Applications}

In this section we compute the shear diffusion constant for a class of
supergravity backgrounds realizing supergravity duals to four dimensional 
gauge theories with eight or less supercharges, namely the
$\caln=2^*$ Pilch-Warner (PW) solution \cite{pw}, the supergravity dual
to the Klebanov-Tseytlin cascading gauge theory (KT) \cite{kt}, and the
$\caln=1$ Maldacena-Nunez solution \cite{mn}.  In all cases we find that
the KSS bound, \eqref{bound}, is saturated.  We then prove that it is
always saturated in the supergravity approximation to gauge theories at
strong 't~Hooft coupling.

\subsection{$\caln=2^*$ gauge theory}
The supergravity dual to $\caln=2^*$ $SU(N)$ gauge theory 
has been proposed in \cite{pw}. The nonextremal deformation 
of the PW solution has been studied in \cite{n2}.  This
solution realizes the supergravity dual to $\caln=4$, $SU(N)$ 
gauge theory softly broken to $\caln=2$. The high temperature 
thermodynamics of this system thus involves a small parameter 
$\xi\equiv m/T$, where $m$ is the mass of the $\caln=2$ hypermultiplet, 
giving rise to  $\caln=4\to\caln=2$ partial supersymmetry breaking.    
That the KSS bound is saturated in this system to leading order in 
$\xi$ has been observed in \cite{kss}. Here, we show that it
is actually valid for arbitrary $\xi$. We will use an exact 
five-dimensional description of the black hole, although since its 10-d
lift is known \cite{n2}, the same result can be obtained directly in 
ten dimensions. 

The relevant near-extremal 5-d Einstein frame metric involves two
functions $A$, $B$ of a radial coordinate $r$
\begin{equation}
ds_5^2=e^{2A}\left(e^{2B}dX_0^2+d\vec{x}\,^2\right)+dr^2\,.
\eqlabel{n2m}
\end{equation}
The horizon is taken to be at $r_{hor}=0$.
One of the equations of motion is (Eq.~(3.19) of \cite{n2})
\begin{equation}
\ln B'+4A+B=4\a+\ln\dd\,,
\eqlabel{constn2}
\end{equation} 
where $\{\a,\dd\}$ are constants specified by the near-horizon 
asymptotics of $A$, $B$ (eq.~(3.20) of \cite{n2})
\begin{equation}
\begin{split}
&A\to \a\,,\\
&e^B\to \dd\ r\,.
\end{split}
\eqlabel{asymp}
\end{equation} 
Notice that we can rewrite \eqref{constn2} as 
\begin{equation}
\left(e^{2B}\right)'=2\dd e^{4\a}\ e^{-4A+B}\,.
\eqlabel{cn22}
\end{equation}

Now we can use \eqref{diff} to compute the shear diffusion constant
\begin{equation}
\begin{split}
\cald_{PW}=&e^{3A}\bigg|_{horizon}\ \int_{0}^{+\infty}dr\ e^{-4A+B}\\
=&e^{3A}\bigg|_{horizon}\ \int_{0}^{+\infty}d\left(e^{2B}\right)\
\frac 12 \dd^{-1}e^{-4\a}\\
=&\frac 12 \dd^{-1}\ e^{-\a}\,.
\end{split}
\eqlabel{dn2}
\end{equation}
Since the black hole temperature is \cite{n2} 
\begin{equation}
T=\frac{1}{2\pi} \dd\ e^{\a}\,,
\eqlabel{tn2}
\end{equation}
we conclude that
\begin{equation}
\cald_{PW}=\frac{1}{4\pi T}\,,
\end{equation}
which generalizes the result of \cite{kss} to all orders in $\xi$.

\subsection{KT gauge theory}
The supergravity dual to $\caln=1$ cascading  $SU(N+M)\times SU(N)$
gauge theory has been proposed in \cite{kt,ks}. 
The nonextremal deformations of the KT solution \cite{kt} 
has been studied in \cite{kt1,kt2,kt3}. In this section we 
follow \cite{kt3}.

The relevant near-extremal 10-d  Einstein-frame metric involves
four functions $x$, $y$, $z$, $w$  of a radial coordinate $u$
\begin{equation}
\begin{split}
ds^2_{10E}&=  e^{2z} (e^{-6x} dX_0^2 + e^{2x} d\vec x\,^2)
+ e^{-2z}  ds^2_6\,,\\
ds^2_6 &= e^{10y} du^2 + e^{2y} (dM_5)^2\,,\\
(dM_5)^2 &=  e^{ -8w}  e_{\psi}^2 +  e^{ 2w}
\left(e_{\theta_1}^2+e_{\phi_1}^2 +
e_{\theta_2}^2+e_{\phi_2}^2\right)\,,\\
e_{\psi} &=  \frac{1}{3} (d\psi +  \cos \theta_1 d\phi_1  +  \cos \theta_2
d\phi_2)\,,
\quad  e_{\theta_i}=\frac{1}{\sqrt 6} d\theta_i\,,  \quad  e_{\phi_i}=
\frac{1}{\sqrt 6} \sin\theta_id\phi_i\,.
\end{split}
\end{equation}
Here $X_0$ is the euclidean time and $\vec x$ are the three longitudinal
D3-brane directions. Also, with our choice of the radial coordinate, the
horizon is at $u=+\infty$, and the boundary is at $u=0$.

The system of equations governing the solution is
\begin{equation}
\begin{split}
&x''=0 \,, \qquad  x= a u \,, \qquad   a={\rm const} > 0\,,\\
&10y'' - 8 e^{8y} (6 e^{-2w} - e^{-12 w})   + \P''=0\,, \\
&10w'' - 12 e^{8y} ( e^{-2w} - e^{-12 w})   - \P'' =0\,, \\
&\P''    + e^{-\P + 4z - 4y-4w} (f'^2 -  e^{2 \P + 8 y+8w} P^2)=0\,, \\
&4z'' -  (Q+ 2 P f)^2  e^{8z}
- e^{-\P + 4z - 4y-4w} ( f'^2 +  e^{2 \P + 8 y+8w} P^2) =0\,,\\
&(e^{-\P + 4z - 4y-4w} f')' - P (Q+ 2 P f) e^{8z} =0\,.
\end{split}
\eqlabel{eq1}
\end{equation}
The integration constants are subject to the zero-energy constraint
$T+V=0$, {\it i.e.}
\begin{equation}
\begin{split}
&y'^2   - 2 z'^2  - 5 w'^2 - \frac 18 \P'^2
- \frac 14  e^{-\P +  4z -4y - 4 w } f'^2\\
& -   \  e^{8y} ( 6 e^{-2w} - e^{-12 w} )
 +  \frac 14 e^{\P+  4z + 4y + 4 w } P^2 +
\frac 18  e^{8z} (Q + 2 P f)^2   = 3 a^2\,. 
\end{split}
\eqlabel{bheq}
\end{equation}

Now we can evaluate the shear diffusion constant for the nonextremal 
KT solution. Following \eqref{diff}, we find 
\begin{equation}
\begin{split}
\cald_{KT}=&e^{5y+3x-2z}\bigg|_{horizon}\int_{horizon}^{boundary}du\ e^{-8x}\\
=&e^{5y+3x-2z}\bigg|_{horizon}\int_{horizon}^{boundary}du\ e^{-8a u}\\
=&\frac{1}{8a}\ e^{5y+3x-2z}\bigg|_{horizon}\,.
\end{split}
\eqlabel{dkt}
\end{equation}
The asymptotics of the solution to \eqref{eq1} at the horizon, 
$u\to +\infty$, are \cite{kt3}
\begin{equation}
\begin{split}
z&\to -a u +z_*\,,\\
y&\to -a u +y_*\,,\\
x&= a u\,.
\end{split}
\eqlabel{horass}
\end{equation}
Thus \eqref{dkt} evaluates to 
\begin{equation}
\cald_{KT}=\frac {1}{8a} e^{5 y_*-2 z_*}\,.
\eqlabel{dktf}
\end{equation}
Since the black hole temperature is \cite{kt3}
\begin{equation}
T=\frac{2}{\pi}a e^{2 z_*-5 y_*}\,,
\eqlabel{tkt}
\end{equation}
we indeed find
\begin{equation}
\cald_{KT}=\frac{1}{4\pi T}\,.
\eqlabel{dktff}
\end{equation}

\subsection{MN gauge theory}
The supergravity dual to $\caln=1$ $SU(N)$ supersymmetric 
Yang-Mills theory has been proposed in \cite{mn}. The nonextremal 
deformation of the MN solution has been studied in \cite{mnbh,gtv}.
In this section we follow \cite{mnbh}.
The relevant near-extremal 10-d  Einstein-frame metric
is
\begin{equation}
\begin{split} 
ds^2_{E}=&c_1(r)^2\left[\t_1^2 dX_0^2+d\vec{x}\,^2\right]+
c_1(r)^2 a^2\Biggl[\frac{dr^2}{ \t_2^2 r^2}+\frac{h(r)}{ 4}\left(
d\theta_1^2+\sin^2\theta_1\ d\phi^2_1\right)\\
&+\frac 14\left(
d\theta_2^2+\sin^2\theta_2\ d\phi^2_2\right)+\frac 14
\left(d\psi+\sum_{i=1}^2\ \cos\theta_i\ d\phi_i\right)^2\
\Biggr]\,,\cr
\end{split}
\eqlabel{mnm} 
\end{equation}

One of the equations of motion is (eq.~(5.55) of \cite{mnbh})
\begin{equation}
\t_1' \t_2=\frac{A}{ c_1(r)^8 h(r) r}\,,
\eqlabel{const}
\end{equation}
where (a constant)   $A$ is the nonextremality parameter. 
Now we can compute 
\begin{equation}
\begin{split}
\cald_{MN}=&a^5 h c_1^8\bigg|_{horizon}\ \int_{horizon}^{boundary} 
dr\ \frac{\t_1}{\t_2 c_1^8 h r}\\
=&a^5 h c_1^8\bigg|_{horizon}\ \int_{horizon}^{boundary} \frac{1}{2 A
a^4}\ 
d\left[\t_1^2\right]\\
=&a^5 h c_1^8\bigg|_{horizon}\ \frac{1}{2 A a^4}\,,
\end{split}
\eqlabel{dmn}
\end{equation}
where in the second line in \eqref{dmn} we used \eqref{const}, and in the
last line we used the boundary conditions $\t_1|_{horizon}=0$ and
$\t_1|_{boundary}=1$.

{}From the near horizon behavior of the metric (eq.~(5.61) of \cite{mnbh})
\begin{equation}
\begin{split}
ds^2_{E}\approx&c_1(r_h)^2\left[\eta^2 dX_0^2+d\vec{x}\,^2\right]
+\frac{c_1^{18}(r_h) a^2 h^2(r_h)}{A^2}\ d\eta^2\\
&+c_1(r_h)^2 a^2\Biggl[\frac {h(r_h)}{4}\left(
d\theta_1^2+\sin^2\theta_1\ d\phi^2_1\right)\\
&+\frac 14\left(
d\theta_2^2+\sin^2\theta_2\ d\phi^2_2\right)+\frac 14
\left(d\psi+\sum_{i=1}^2\ \cos\theta_i\ d\phi_i\right)^2\
\Biggr]\,,
\end{split}
\eqlabel{nhmnm}
\end{equation}
we find 
\begin{equation}
\frac 1T= 2\pi\ \frac{c_1^8 a h}{A}\bigg|_{horizon}\,.
\eqlabel{tmn}
\end{equation}
Comparing \eqref{dmn} and \eqref{tmn} we find
\begin{equation}
\cald_{MN}=\frac {1}{4\pi T}\,.
\eqlabel{dtmn}
\end{equation}
Hence for at three non-trivial supergravity backgrounds we have found
a common result, that $\cald=1/4\pi T$.  In all cases, we have taken
advantage of reducing the $u$ integral in \eqref{diff} to a boundary term.
This suggests that saturation of the KSS bound is in fact universal,
depending only on the thermal nature of the background spacetime.  We
now examine this in detail.

\subsection{General supergravity backgrounds}

Our key observation is that before turning on any nonextremality, the
Poincare symmetry of the background geometry guarantees that 
\begin{equation}
R_{tt}+R_{xx}=0\,,
\eqlabel{constr}
\end{equation} 
where $R_{\mu\nu}$ is the Ricci tensor in an orthonormal frame. 
Clearly, an analogous condition must be satisfied for the full 
stress tensor of the matter supporting the geometry
\begin{equation}
T_{tt}+T_{xx}=0\,.
\eqlabel{constt}
\end{equation} 
Because turning on the nonextremality will not modify \eqref{constt}, we
see that \eqref{constr} is valid away from extremality as well.  We now
show that \eqref{constr} is sufficient to explicitly evaluate \eqref{diff},
yielding the result 
\begin{equation}
\cald T=\frac{1}{4\pi}\,.
\eqlabel{dgen}
\end{equation}

Consider the following $D=d+p+q$ dimensional background
\begin{equation}
ds_D^2=\om_1^2(y) \left(g_{\mu\nu}(x)dx^\mu dx^\nu\right)
+\om_2^2(y)\left(\hg(z)_{\a\b}dz^\a dz^\b+\tg(y)_{mn}dy^m dy^n\right)\,,
\eqlabel{metricD}
\end{equation}
where $g_{\mu\nu}$ is $d$-dimensional, $\hg_{\a\b}$ is $p$-dimensional and 
$g_{mn}$ is $q$-dimensional.  An explicit computation yields
\begin{equation}
\begin{split}
R_{\mu\nu}&=r_{\mu\nu}-g_{\mu\nu}\biggl(\om_2^{-2}\om_1\n^2\om_1
+(d-1)\om_2^{-2}\left(\n\om_1\right)^2+(D-d-2)\om_2^{-3}\om_1\n\om_1\n\om_2
\biggr)\,,\\
R_{\a\b}&=\hr_{\a\b}-\hg_{\a\b}\biggl(\om_2^{-1}\n^2\om_2
+(D-d-3)\om_2^{-2}\left(\n\om_2\right)^2+d\om_2^{-1}\om_1^{-1}\n\om_1\n\om_2
\biggr)\,,
\end{split}
\eqlabel{riccicomp}
\end{equation} 
where $\n$ is with respect to $g_{mn}$, and $r_{\mu\nu},\hr_{\a\b},
\tr_{mn}$ are Ricci tensors computed from $g_{\mu\nu}, \hg_{\a\b}$
and $\tg_{mn}$ correspondingly.

To be relevant for the nonextremal RG flows of 4-d gauge theories,
we now take $d=1$, $p=3$ and $D=10$.  Also, we have
$r_{\mu\nu}=\hr_{\a\b}=0$, and $\om_1(r)=\om_2(r)\t(r)$ 
depends only on the radial coordinate of $\tg_{mn}$. 
$\t(r)$ is the nonextremality warp factor with the boundary 
conditions
\begin{equation}
\t(r)\bigg|_{r=r_0}=0,\qquad \t(r)\bigg|_{r=\infty}=1\,,
\eqlabel{bc}
\end{equation}
where we take the horizon to be at $r=r_0$ and the boundary to be
at $r=\infty$.
Given \eqref{riccicomp}, the linear combination of Ricci components,
\eqref{constr}, gives rise to
\begin{equation}
\n^2\t+8\om_2^{-1}\n\om_2\n\t=0\,.
\eqlabel{teq}
\end{equation}
Assuming the radial dependence as above, we find 
the first integral of \eqref{teq} to be
\begin{equation}
\frac {d\t}{dr}\ \tg_{rr}^{-1/2}\tg_5^{1/2}=A \om_2^{-8}\,,
\eqlabel{1int}
\end{equation}
where $A$ is an integration constant, related to the 
temperature as we explain below.  Here we have decomposed the 
$q=6$ dimensional metric $\tg_{mn}$ as follows
\begin{equation}
g_{mn}(y) dy^mdy^n=\tg_{rr} dr^2+\tg_{5\,ij}dy^idy^j\,.
\eqlabel{decomtg}
\end{equation}

It is easy to see now that the expression \eqref{diff} reduces to
\begin{equation}
\begin{split}
\cald=&\frac{\sqrt{-G(r)}}{\sqrt{-G_{tt}(r)G_{rr}(r)}}\bigg|_{horizon}
\int_{r_0}^{\infty}\frac{\t d\t}{A}\\
=&\frac{\sqrt{-G(r)}}{\sqrt{-G_{tt}(r)G_{rr}(r)}}\bigg|_{horizon}
\frac{1}{2A}\\
=&\frac{1}{2A}\ \om_2^8(r_0)\tg_5^{1/2}(r_0)\,.
\end{split}
\eqlabel{difgen}
\end{equation}
Furthermore, note that using \eqref{1int} near the horizon yields
\begin{equation}
ds_{10}^2=-\left(dt\right)^2\om_2^2(r_0)\t^2+
\om_2^2(r_0)\ g_{rr}(r_0)\left(\frac{d\t}{A}\right)^2 g_{rr}^{-1}(r_0)
\tg_5(r_0)\om_2^{16}(r_0)\,,
\end{equation}
from which we can read off the temperature
\begin{equation}
T=\frac{A}{2\pi\tg_5(r_0)^{1/2}\om_2^8(r_0)}\,.
\eqlabel{tfin}
\end{equation} 
Combining \eqref{difgen} and \eqref{tfin}, we the arrive at
\begin{equation}
\cald=\frac{1}{4\pi T}\,.
\eqlabel{satur}
\end{equation}
This proves our claim that the KSS bound is always saturated in the
supergravity dual, at least to this leading order in corrections.

\section{A comment on $\a'$ corrections}

The explicit examples in previous section, and the general theorem 
in section 3.4, demonstrate that the KSS bound, \eqref{bound}, is
saturated in all supergravity backgrounds realizing holographic duals
to gauge theories at large (strictly speaking infinite) 't~Hooft coupling
$\lambda\equiv g_{YM}^2 N$. 
On the other hand, as pointed out in \cite{kss}, 
for typical matter (\ie, water under normal conditions)
$4\pi \cald T\gg 1$. These two observations together make us 
conjecture that 
\begin{equation}
\cald=\frac{f(\lambda)}{4\pi T}\,,
\eqlabel{conj}
\end{equation}
where $f(\lambda)$ is (in principle) a computable function of the
't~Hooft coupling, such that for arbitrary $\lambda$,
\begin{equation}
f(\lambda)\ge 1\,,
\eqlabel{fl1}
\end{equation}
and 
\begin{equation}
f(\lambda)\to 1_+, \qquad \lambda\to \infty\,.
\eqlabel{fl2}
\end{equation}

Since on the supergravity side of the gauge/string correspondence,
't~Hooft coupling corrections translate into string theory
$\a'$-corrections, verification of the conjecture \eqref{fl2} 
would involve the study of $\a'$ corrections to the hydrodynamics.
Realistically, this can be done for the near-extremal D3-branes. 
In fact, using the KSS expression for the diffusion coefficient,
\eqref{diff}, applied to the $\a'$-corrected metric of the 
near-extremal D3-branes \cite{gkt}, we found that the bound \eqref{bound}
is {\it violated}
\begin{equation}
\cald T=\frac{1}{4\pi}\biggl(1-15\gamma+\calo(\gamma^2)\biggr)\,,
\eqlabel{violation}
\end{equation} 
where 
\begin{equation}
\gamma=\frac 18\xi(3)\a'^3\,.
\eqlabel{gammadef}
\end{equation}
We would like to emphasize, however, that this result assumes that the
dispersion relation for the low-energy gravitational shear perturbations
(which led to the diffusion coefficient expression \eqref{diff}) is not
modified by the $\a'$ corrections. This assumption is very likely incorrect,
and one thus has to do complete analysis of the metric fluctuations
themselves. We hope to return to this problem in the future.

\section*{Acknowledgments}

We would like to thank Joe Polchinski and Arkady Tseytlin for interesting
discussions.  We are particularly thankful to Andrei Starinets for 
bringing the problem to our attention and for numerous discussions and
explanations.  AB would like to thank KITP for hospitality 
while this work was done. This research was supported in part by the
National Science Foundation under Grant No.~PHY99-07949 (AB) and the
US Department of Energy under Grant No.~DE-FG02-95ER40899 (JTL).


\end{document}